\documentclass[elsarticle]{revtex4}
\usepackage{graphicx}
\usepackage{subfig}
\begin{document}
\voffset 1.8cm

\title{How good are the Garvey-Kelson predictions of nuclear masses?}

\author{Irving O.~Morales} 
\affiliation{Instituto de Ciencias Nucleares,
Universidad Nacional Aut\'{o}noma de M\'{e}xico, Apartado Postal
70-543, 04510 M\'{e}xico, D.F., Mexico} 

\author{J.C.~L\'{o}pez Vieyra}
\affiliation{Instituto de Ciencias Nucleares,
Universidad Nacional Aut\'{o}noma de M\'{e}xico, Apartado Postal
70-543, 04510 M\'{e}xico, D.F., Mexico} 

\author{J.G.~Hirsch}
\affiliation{Instituto de Ciencias Nucleares,
Universidad Nacional Aut\'{o}noma de M\'{e}xico, Apartado Postal
70-543, 04510 M\'{e}xico, D.F., Mexico}

\author{A.~Frank}
\affiliation{Instituto de Ciencias Nucleares,
Universidad Nacional Aut\'{o}noma de M\'{e}xico, Apartado Postal
70-543, 04510 M\'{e}xico, D.F., Mexico} 

\date{\today}

\begin{abstract}
The Garvey-Kelson relations are used in an iterative process to predict nuclear masses in the neighborhood of nuclei with measured masses.  Average errors in the predicted masses for the first three iteration shells are smaller than those obtained with the best nuclear mass models. Their quality is comparable with the Audi-Wapstra extrapolations, offering a simple and reproducible procedure for short range mass predictions. A systematic study of the way the error grows as a function of the iteration and the distance to the known masses region, shows that a correlation exists between the error and the residual neutron-proton interaction, produced mainly by the implicit assumption that $V_{np}$ varies smoothly along the nuclear landscape.\\  
\end{abstract}

\maketitle

\section{Introduction.}
Mass is the most basic property of atomic nuclei. Knowledge of nuclear masses is essential not only for nuclear physics but for other branches of physics.
In astrophysics they are required as a basic input for the calculation of diverse processes in the evolution of stars \cite{Rod88}. 
Several methods and models have been developed to predict the unknown nuclear masses \cite{Lun03}, all of them trying to use the essential physical information determining the experimentally known masses. Unfortunately the predictions of these approaches tend to diverge from each other, so there is a permanent search for better theoretical models that reduce the difference with the experimental masses and produce reliable predictions for unstable nuclei.\\

Besides the global formulas, there are a number of local mass formulas. These local methods are usually effective when calculating the mass of a nucleus, or a set of nuclei, which are fairly close to nuclei of known mass, taking advantage of the relative smoothness of the masses as a function of proton and neutron numbers to deduce systematic trends. Among these methods, there are a set of algebraic relations for nuclear neighbors known as the Garvey-Kelson relations \cite{Gar66}.\\

The surface in the neutron number $N$, proton number $Z$ space of the known nuclear masses, as well as the surface of its ``derivatives'': the two-neutron and two proton separation energies, and the Q$_\alpha$ and Q$_\beta$ values, vary smoothly as a function of $N$ and $Z$. This property is fundamental and permits extrapolation to the unknown region using only the systematic trends of this surface \cite{Bor93}. This method provides the best short range mass extrapolations \cite{Lun03}, which have been published for sets of three or four nuclides in the neighborhood of those with measured masses in the Atomic Mass Evaluations \cite{Aud95,Aud03}.  These predictions are performed nucleus by nucleus, combining a (rather elaborate) graphical analysis with relevant physical information \cite{Bor93}.\\

In the present article it is shown that the well known Garvey-Kelson relations can be used in an iterative way to predict masses as accurately as in AME, and significantly better than any nuclear mass model, for the first three iteration shells, which cover a large number of nuclei. After a short review of the Garvey-Kelson relations, comparisons with several mass models are presented, as well as a study of the systematic behavior of the error in this relations.\\

The Garvey-Kelson relations \cite{Gar66} are algebraic expressions that relate the masses of six neighboring nuclides. These relations can be derived in an (extreme) single-particle shell model by constructing a suitable combination of neighboring masses in such a way that the residual neutron-neutron, proton-proton and neutron-proton interactions cancel out. The main assumption is that the positions of the single particle levels and the residual interactions between nucleons vary smoothly with atomic number. Two algebraic equations can be obtained in this way 
\begin{equation}  \begin{array}{c}
    M(N+2,Z-2)-M(N,Z)+M(N,Z-1)-M(N+1,Z-2)+M(N+1,Z)-M(N+2,Z-1)=0\\ 
  \end{array}
  \label{gk1}
\end{equation}
\begin{equation}  \begin{array}{c}
    M(N+2,Z)-M(N,Z-2)+M(N+1,Z-2)-M(N+2,Z-1)+M(N,Z-1)-M(N+1,Z)=0\\ 
  \end{array}
  \label{gk2}
\end{equation}\\
where $M(N,Z)$ is the nuclear mass. None of the nuclei in a Garvey-Kelson relation should have $N=Z$ odd, and for relation \ref{gk1} to be fulfilled it is necessary that $N > Z$, to secure a proper cancelation of the isospin dependence of the residual interactions\cite{Gar66}. \\ 

For each relation it is possible to obtain the mass of a given nucleus in terms of the remaining five masses involved, which can be done in six different ways. Using the two relations it is possible to obtain up to 12 estimates for the mass of a nucleus \cite{Bar05,Bar08}. To estimate a mass using these relations it is necessary to know at least five of the six masses involved \cite{Bar05,Bar08}. Although only a rather small set of nuclei with unknown masses satisfy this condition, it is possible to make further predictions through an iterative procedure, by employing the masses computed in previous steps. In this article we analize in a precise fashion how the average error behaves as a function of the number of iterations.We show that it is possible to achieve  very reliable mass predictions for the first three iteration steps, which add up to a large number of nuclei.\\

\section{Short range probes.}
In order to compare the predictive power of the Garvey-Kelson process with the Audi-Wapstra extrapolations for the AME95\cite{Aud95} compilation, we start with the experimentally known masses in AME95 as the initial set, predict the new masses reported in the AME03 compilation \cite{Aud03}, and compare them with their measured values.\\

The Garvey-Kelson relations have been used before in an iterative way to predict masses of nuclei not too far from known ones. A study of their predictive power for this test have been published\cite{Lun03}. However, to the best of our knowledge, an iterative process considering the 12 possible estimates for each mass has not been analyzed. Figure \ref{gaudi} shows the number of estimates used to predict each nucleus for the test described before. While most of these nuclei are predicted with only one estimate, it is important to note that it is just one of twelve possibilities. We also observe that many nuclei are predicted with more than one estimate. Given the simplicity of the method, we believe that these two factors make it worthwhile  to make a new analysis of predictability for the Garvey-Kelson relations.\\

\begin{figure}
\begin{center}
\subfloat{\label{numrel}\includegraphics[width=8.0cm]{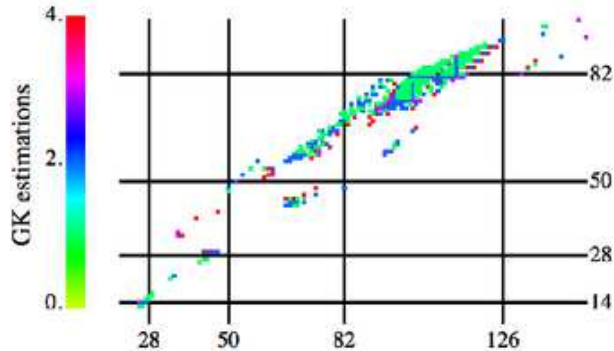}} 
\caption{Number of Garvey-Kelson estimates used to predict each nucleus.}
\label{gaudi}
\end{center}
\end{figure}

At each iteration step we estimate the mass for all the nuclei in the prediction set for which it is possible to use at least one Garvey-Kelson relation. If it is possible to use more than one relation, we take the average value of these estimates. For all the masses computed in this fashion we calculate four relevant quantities at each iteration, the root mean square deviation ($\sigma^{i}$ and $\sigma^{i}_{mod}$) for the nuclei predicted on iteration $i$.

\begin{equation}  
\sigma^{i}=\left(\frac{1}{N_{i}} \sum_{j=1}^{N_{i}}(M^{GK}_{i,j}-M_{i,j}^{exp})^{2}\right)^{1/2}
 \label{gk1}
\end{equation}

\begin{equation}  
\sigma^{i}_{mod}=\left(\frac{1}{\sum w_{i,j}} \sum_{j=1}^{N_{i}}w_{i,j}((M^{GK}_{i,j}-M_{i,j}^{exp})^{2}-(e_{i,j})^{2}\right)^{1/2}
 \label{gk1}
\end{equation}

where the index $j$ goes over the $N_{i}$ nuclei predicted on iteration $i$, $e_{i,j}$ is the experimental error and $w_{j}$ is defined as

\begin{equation}
w_{i,j}=\frac{1}{((e_{i,j})^{2}+(\sigma^{i}_{mod})^{2})^{2}}
\end{equation}

and the accumulated root mean square deviation ($A\sigma^{i}$ and $A\sigma^{i}_{mod}$) for the nuclei predicted up to iteration $n_{i}$

\begin{equation}  
A\sigma^{i}=\left(\frac{1}{\sum N_{i}}\sum_{i=1}^{n_{i}} \sum_{j=1}^{N_{i}}(M^{GK}_{i,j}-M_{i,j}^{exp})^{2}\right)^{1/2}
 \label{gk1}
\end{equation}

\begin{equation}  
A\sigma^{i}_{mod}=\left(\frac{\displaystyle \sum_{i=1}^{n_{i}} \displaystyle \sum_{j=1}^{N_{i}}w_{i,j}((M^{GK}_{i,j}-M_{i,j}^{exp})^{2}-e_{i,j}^{2}}{\sum \sum w_{i,j}}\right)^{1/2}
 \label{gk1}
\end{equation}

where the \emph{mod} versions of $\sigma$ and $A\sigma$ are the \emph{model standard deviation} \cite{Moll88}. The predicted masses are then added to the ``known" set for the next iteration.\\

Table \ref{audi} and Figure \ref{gaudi} allow a comparison, at each iteration, between the $\sigma^{i}$ and $A\sigma^{i}$ mass deviation for Garvey-Kelson and Audi-Wapstra.

 In the second column of Table  \ref{audi} the number of nuclei whose masses are predicted up to each iteration is given. Between parenthesis is the number of nuclei predicted at each iteration. From the 389 new masses, 93 of them are predicted in the first iteration, and 198 in the first three iterations. We see that for the first three iterations the Garvey-Kelson relations provide mass predictions which are comparable to those  computed in Ref. \cite{Aud95}. The quality of these predictions is significantly better than those provided by the best global models \cite{Lun03}.\\ 

It is interesting to compare this predictions with the previous ones\cite{Lun03}. In the cited reference they mention "in the limit of seven iterations, for example, 242 out of the 382 new masses could be calculated with deviation $A\sigma_{mod}=0.232$ MeV. A total of 21 iterations were required to reach 340 of these masses, but with the resulting large rms deviation of  $A\sigma_{mod}=0.717$ MeV". Following the iterative process used in this paper, 242 nuclei are predicted after only 5 iterations and are predicted with an $A\sigma^{i}_{mod}$ about 0.35 MeV.  To reach 340 predicted nuclei only 11 iterations are needed and the deviation with the experimental values is $A\sigma^{i}_{mod}=0.628$. The inclusion of the twelve possible estimates allows the iterative process to reach further on each iteration and because of this the error grows more slowly.\\

\begin{figure}[h]
\begin{center}
\subfloat{\label{95pos}\includegraphics[width=8cm]{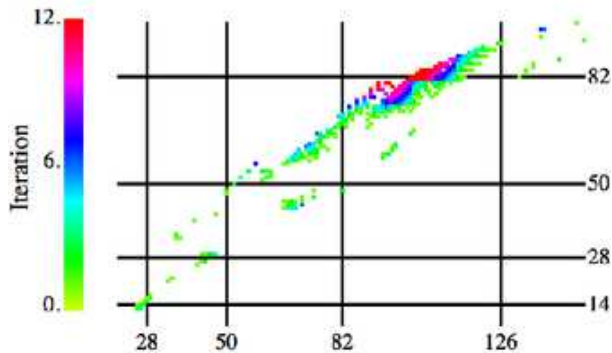}}
\caption{Set of nuclei whose masses are predicted on each iteration for the AME95-03 probe.}
\label{posit}
\end{center}
\end{figure}

The Garvey-Kelson iterative process is a simple, systematic and reliable option to estimate nuclear masses which doesn't require an elaborate graphical analysis, in the vicinity of the experimentally known region it is competitive with Audi-Wapstra extrapolations. After the fourth iteration the iterative process tends to increase rapidly and its predictions are no longer reliable. The color code in Figure \ref{95pos} shows the nuclei predicted at each iteration step for this test. It would be useful to develop new strategies to allow the Garvey-Kelson relations to reach further.\\

\begin{figure}
\begin{center}
\subfloat{\label{95gkaurms}\includegraphics[width=8cm]{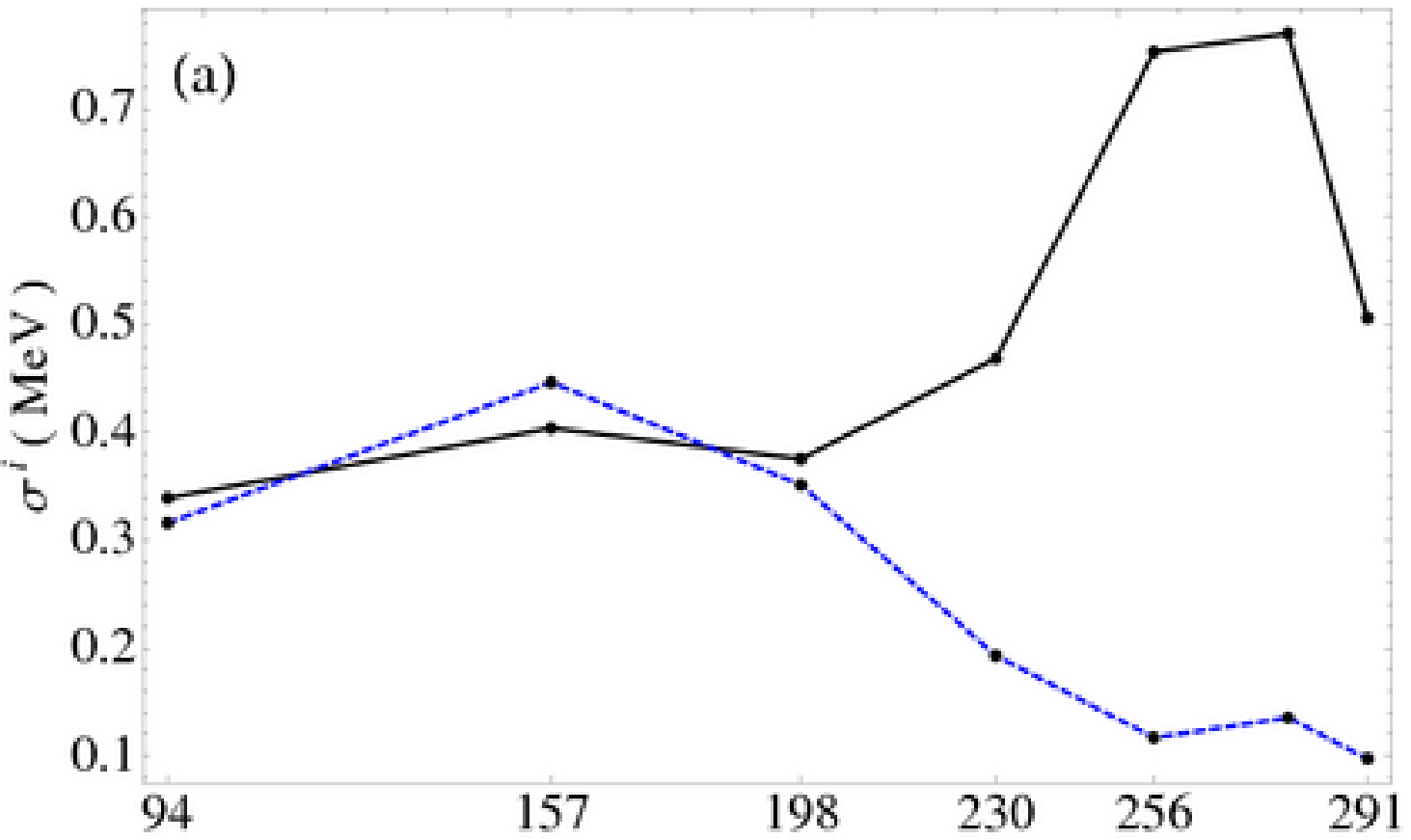}}\qquad
\subfloat{\label{95auarms}\includegraphics[width=8cm]{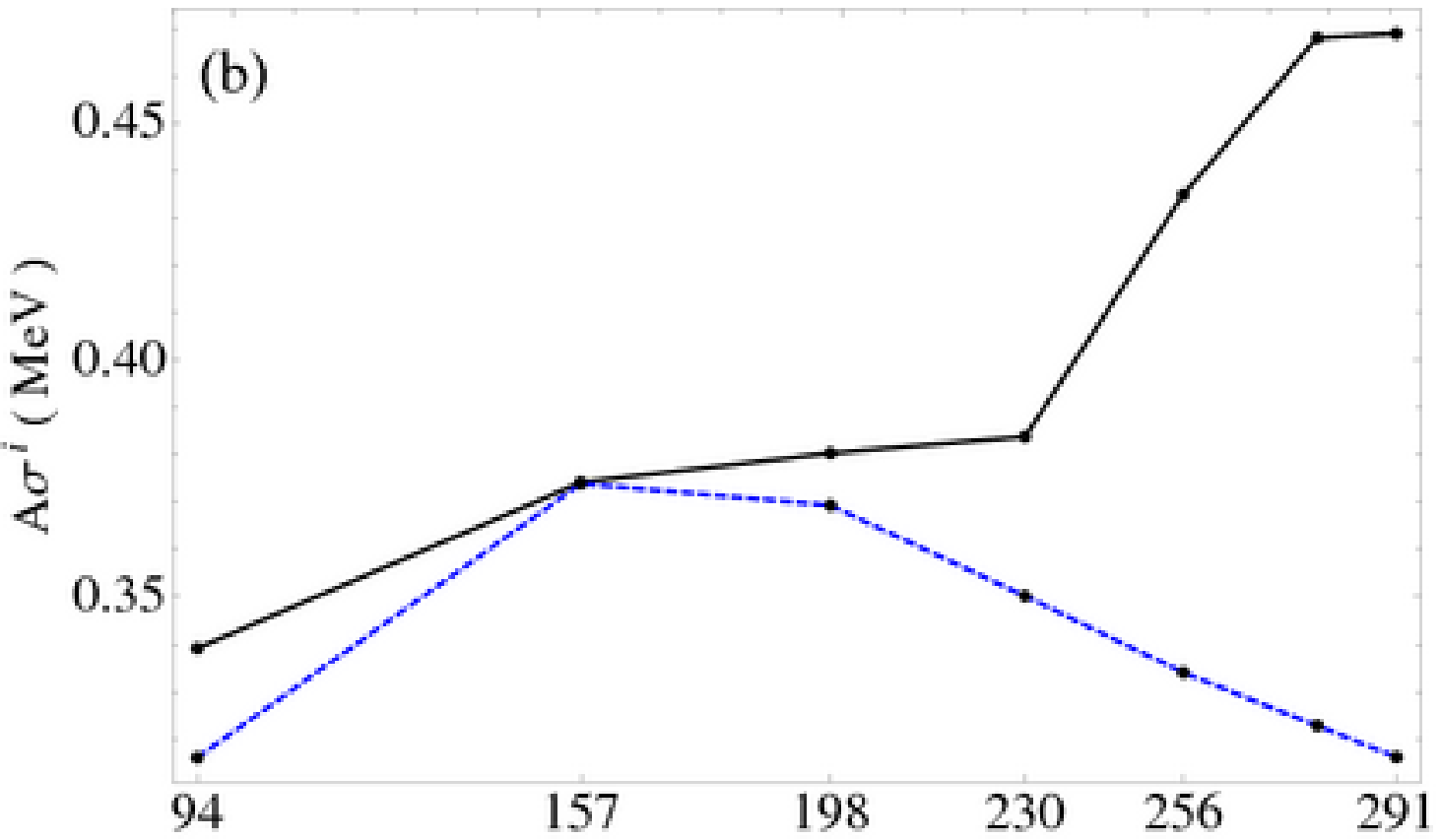}}
\caption{Comparison between the $\sigma^{i}$ (a) and $A\sigma^{i}$ (b) deviation in MeV for the predictions of the nuclei on each iteration for the Garvey-Kelson iterative process (black) and the Audi-Wapstra extrapolations (blue) as a function of the accumulated number of nuclei with predicted masses.}
\label{gaudi}
\end{center}
\end{figure}

Fourteen different tests have been developed in order to gauge the predictive power of different mass models \cite{Tem08}. One of them, the 95-03 probe, is particularly well suited to test short range predictions and will be used to compare the ability of the Garvey-Kelson relations to predict masses generated by some of representative global models. Starting with the AME03 data set \cite{Aud03}, we divide it into two subsets, one assumed to be known, while the other is predicted.  We analyze the cases with $N\geq8,Z\geq8$ and $N\geq28,Z\geq28$.\\

The Garvey-Kelson predictions do not vary if we consider $N\geq8,Z\geq8$ or $N\geq28,Z\geq28$ because of its local character.  However, some of the mass models improve their predictions when we do not include the light nuclei. In other words, the $\sigma^{i}$ curve as a function of distance oscillates less if the light nuclei are not present in the fit.\\

\begin{figure}[h]
\begin{center}
\subfloat{\label{95rms}\includegraphics[width=8cm]{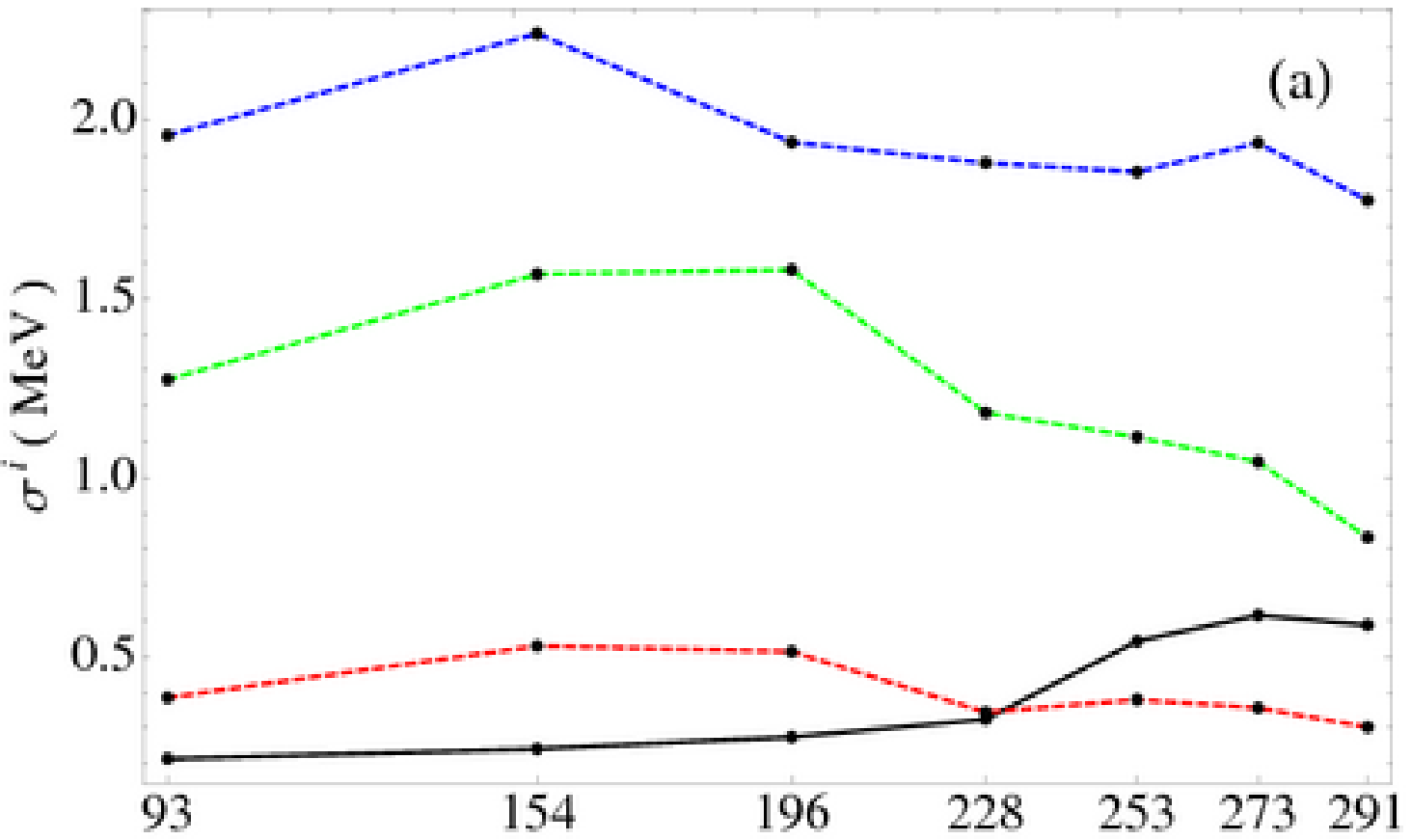}} \qquad
\subfloat{\label{95arms}\includegraphics[width=8cm]{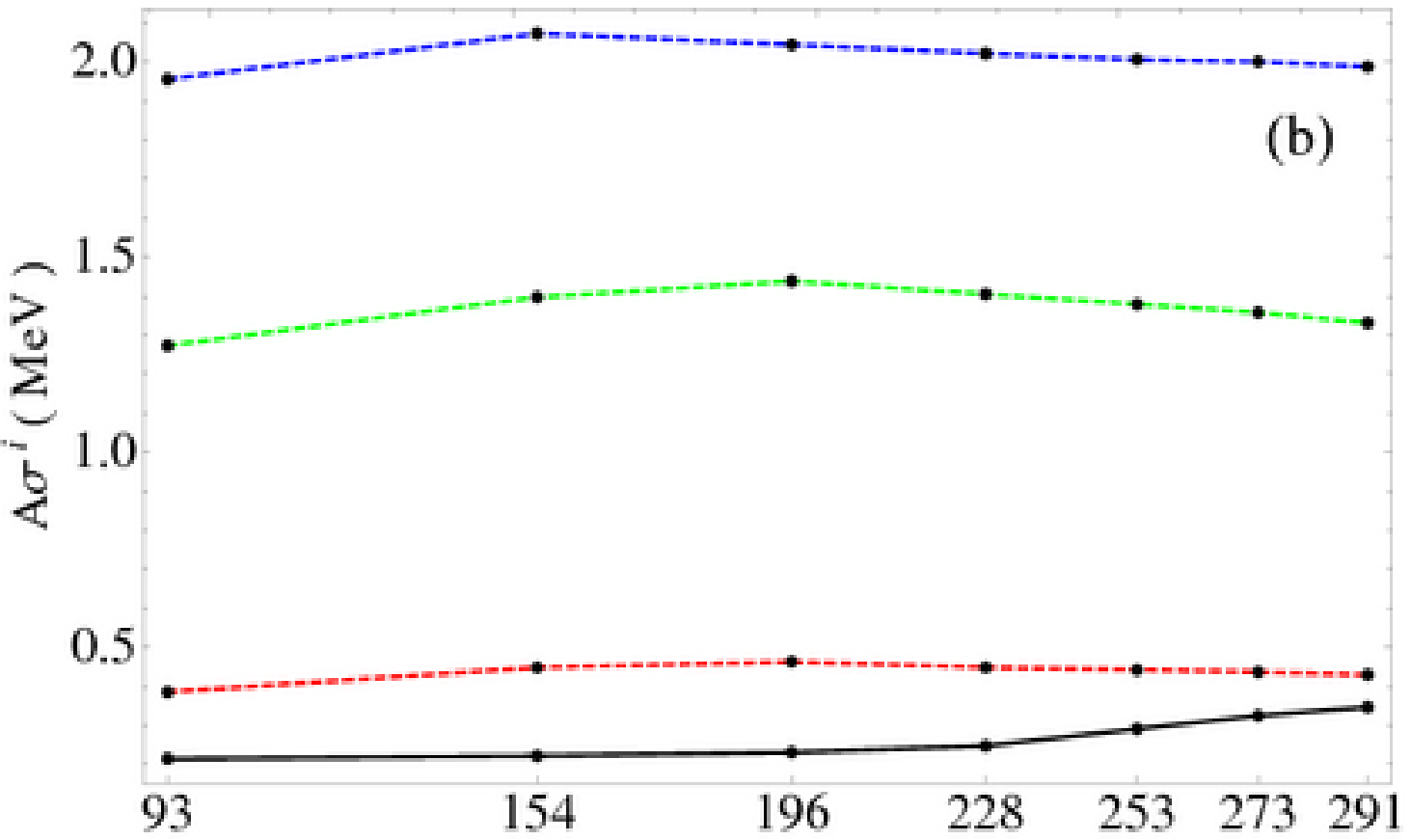}}
\caption{$\sigma^{i}$ (a) and $A\sigma^{i}$ (b) deviation (in MeV) for the mass predictions on each iteration, for the AME95-03 probe with $N\geq8,Z\geq8$, of the Garvey-Kelson iterative procedure (black), LDM(blue dotted),  LDMM(green dotted), DZ(red dotted).}
\label{game9503}
\end{center}
\end{figure}

\begin{figure}[h]
\begin{center}
\subfloat{\label{9528rms}\includegraphics[width=8cm]{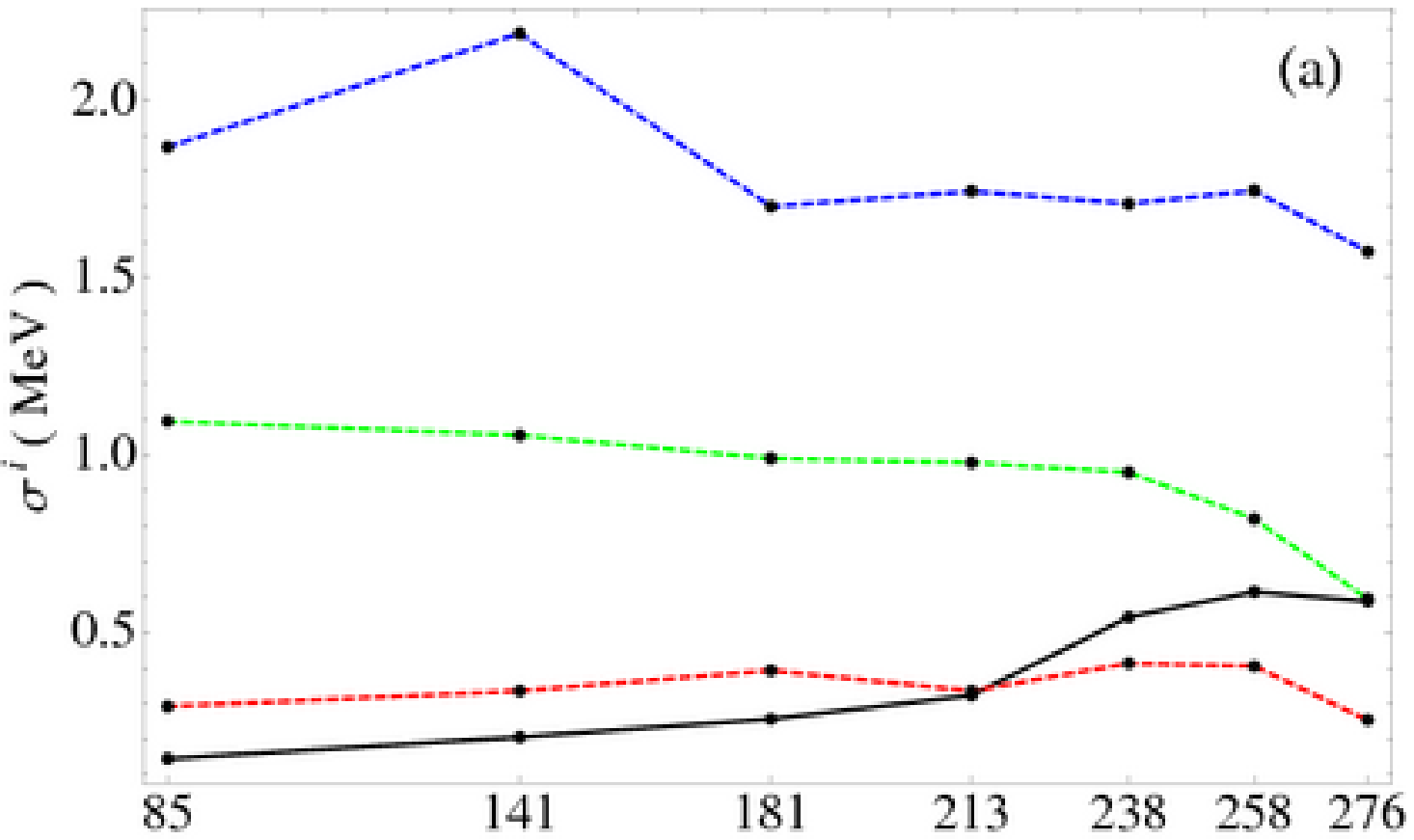}} \qquad
\subfloat{\label{9528arms}\includegraphics[width=8cm]{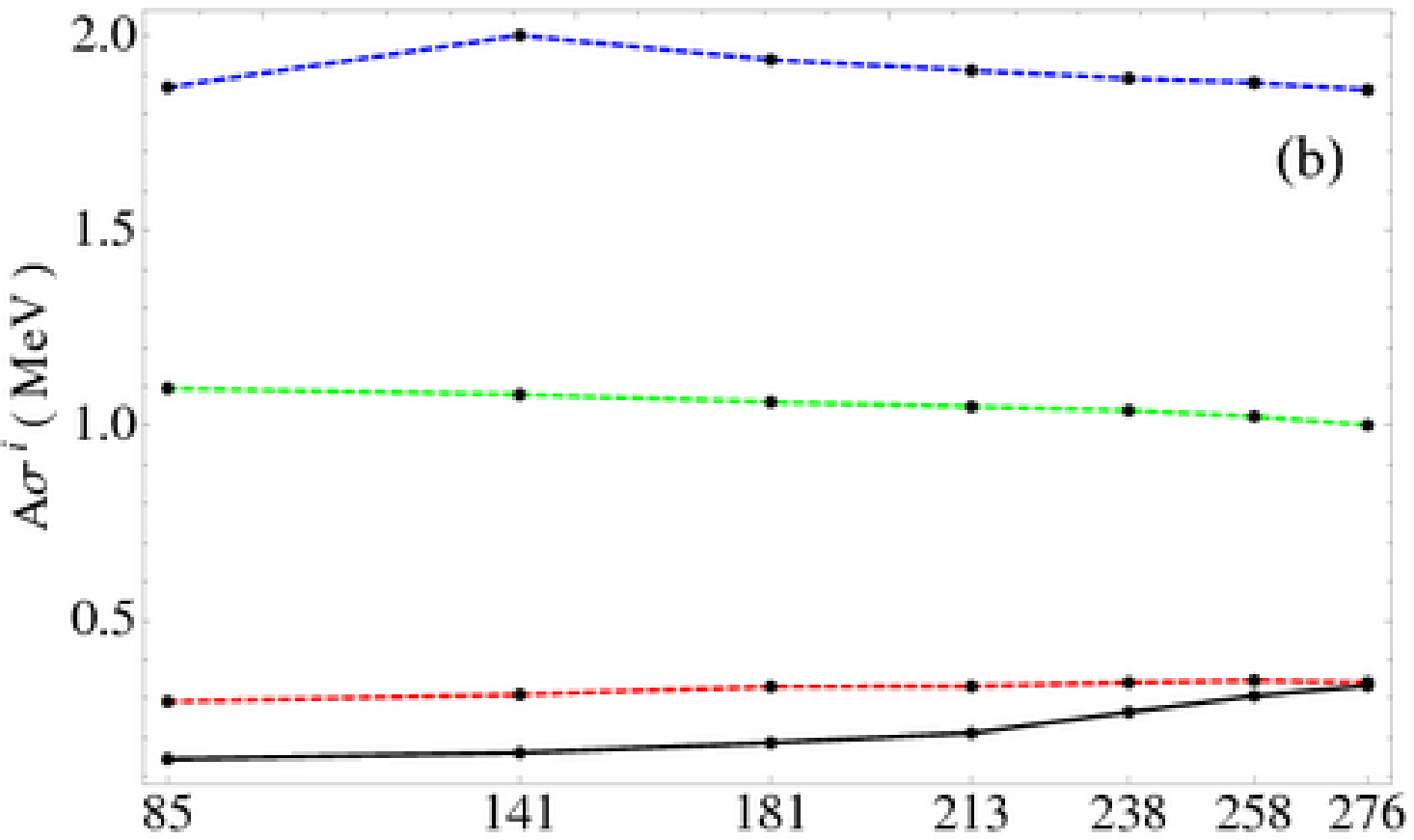}}
\caption{The same as Fig. \ref{game9503} for $N\geq 28,Z\geq 28$ .}
\label{g28ame9503}
\end{center}
\end{figure}

For the 95-03 test with $N\geq8,Z\geq8$ there are 1760 nuclei  in the initial (fitting) subset, while the prediction consists of 389 nuclei. The case with $N\geq28,Z\geq28$ has in the initial subset 1454 nuclei and 371 in the prediction set.\\

The performance of the Garvey-Kelson relations (GK) to predict masses will be measured against the following three mass models \cite{Tem08}: the Bethe-Weizsacker Liquid Drop Model (LDM), modified to include volume and surface contributions in the symmetry energy and a Wigner correction \cite{Dan03,Die07}; the same LDM including schematic shell corrections, linear and quadratic in the number of valence nucleons (LDMM) \cite{Die07,Men08} and the Duflo-Zuker (DZ) model \cite{Duf95} with 33 parameters. The LDMM and the DZ models contain information about the single-particle (shell) behavior of nuclei through microscopic corrections, while the LDM is purely macroscopic. In each model the parameters are fitted to the nuclei in the initial set for each probe. The model then produce a prediction, which can be compared to the corresponding GK prediction. The comparison with the Duflo-Zuker model is especially interesting since it produces the lowest errors against measured masses in g
 lobal calculations.\\

Figure \ref{game9503} shows the $\sigma^{i}$ and $A\sigma^{i}$ deviations in MeV for the predicted nuclei on each iteration for the three models in the case $N\geq8,Z\geq8$. It is evident that the Garvey-Kelson procedure is at least as good as the models, including microscopic corrections, LDMM and DZ, up to iteration 7 and is better than the LDM up to iteration 13. We can also observe that the predictions of the microscopic models are quite stable through all the iterations, especially for the Duflo-Zucker model. Figure \ref{g28ame9503} shows the $\sigma^{i}$ and $A\sigma^{i}$ comparison for the case with $N\geq28,Z\geq28$. Again the Garvey-Kelson procedure improves the results up to 7 and 13 iterations, respectively, when comparing with the precision of the LDMM and DZ models and with the LDM, respectively.\\

\section{Long range probes.}
It has been demonstrated previously\cite{Lun03} and confirmed systematically in this work, that the Garvey-Kelson relations are an excellent tool to predict nuclear masses in the vicinity of the experimentally known region. However,  the rms deviation grows rapidly because of the use of previously predicted masses on each new iteration. It is important to develop strategies to make reliable predictions further away from known masses.  We now present a systematic analysis of the GK errors and their patterns of increase. \\

\begin{figure}
\begin{center}
\subfloat{\label{numrel}\includegraphics[width=8.0cm]{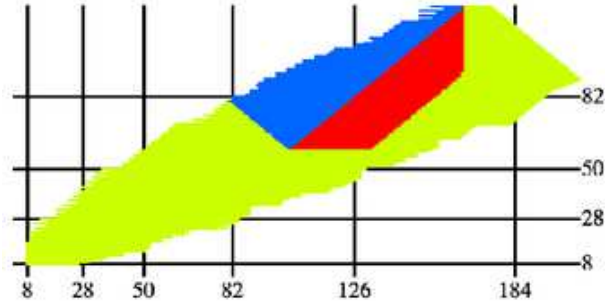}} 
\caption{Nuclei used as input (Blue) and those who were predicted (Red) using the Garvey-Kelson relations iterative procedure in the long range probe.}
\label{poslot}
\end{center}
\end{figure}

To analyze the error produced on each iteration the AME95-03 probe is not optimal, We require a test covering a larger extent of nuclei, as it is very important to control the position of the predicted nuclei on each iteration.  This ensures that the iterations truly correspond to distance to the known region. Each of the predictions of previously used models (LDM, LDMM and DZ) is a good set of data consisting of approximately 9000 binding energies that satisfy the Garvey-Kelson relations to a good approximation. Using these sets we construct the long range probe, again dividing the data into two subsets. All nuclei with $A>160$ and $(N-Z)<44$ are used as known masses and are shown in figure \ref{poslot} in blue. Nuclei with  $A>160$ and $(N-Z)\geq44$ are the ones that will be predicted by the iterative procedure and are shown in red.\\

\begin{figure}[h]
\begin{center}
\subfloat{\label{iterlot}\includegraphics[width=8cm]{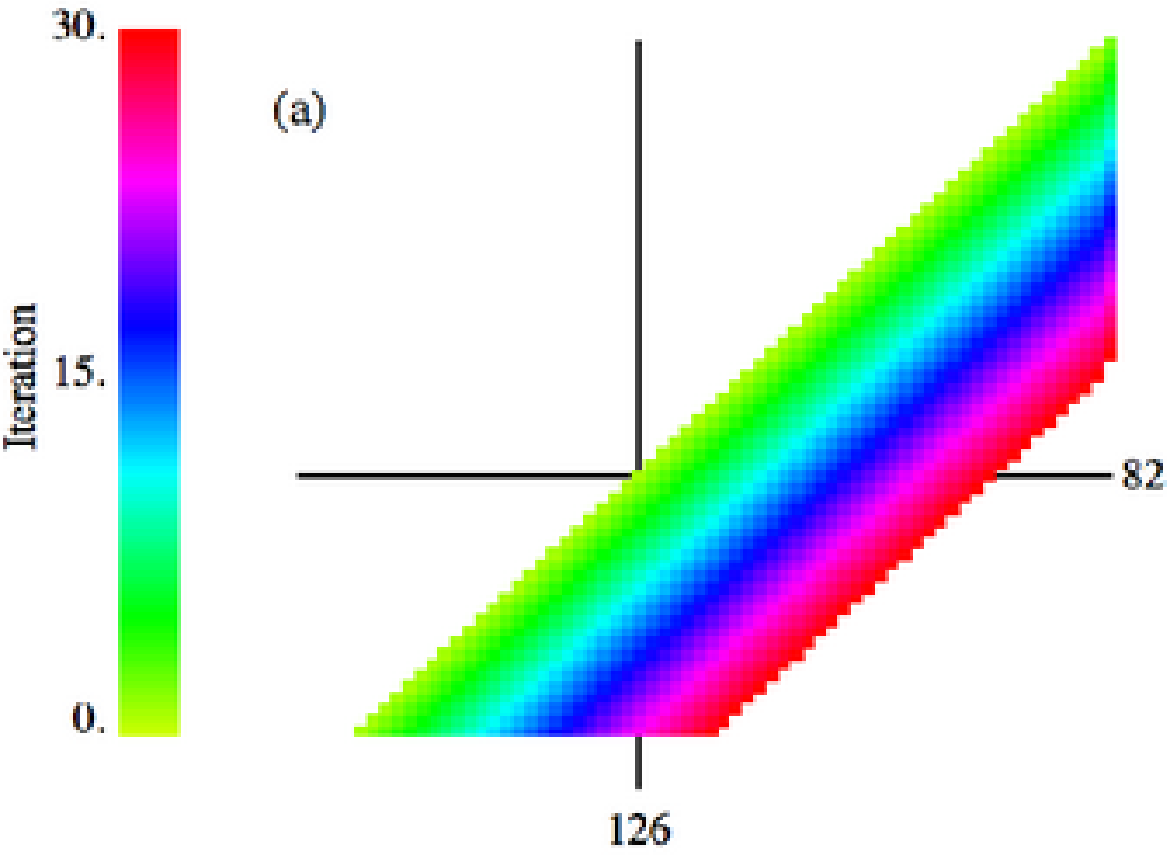}}\qquad
\subfloat{\label{rmsac}\includegraphics[width=8cm]{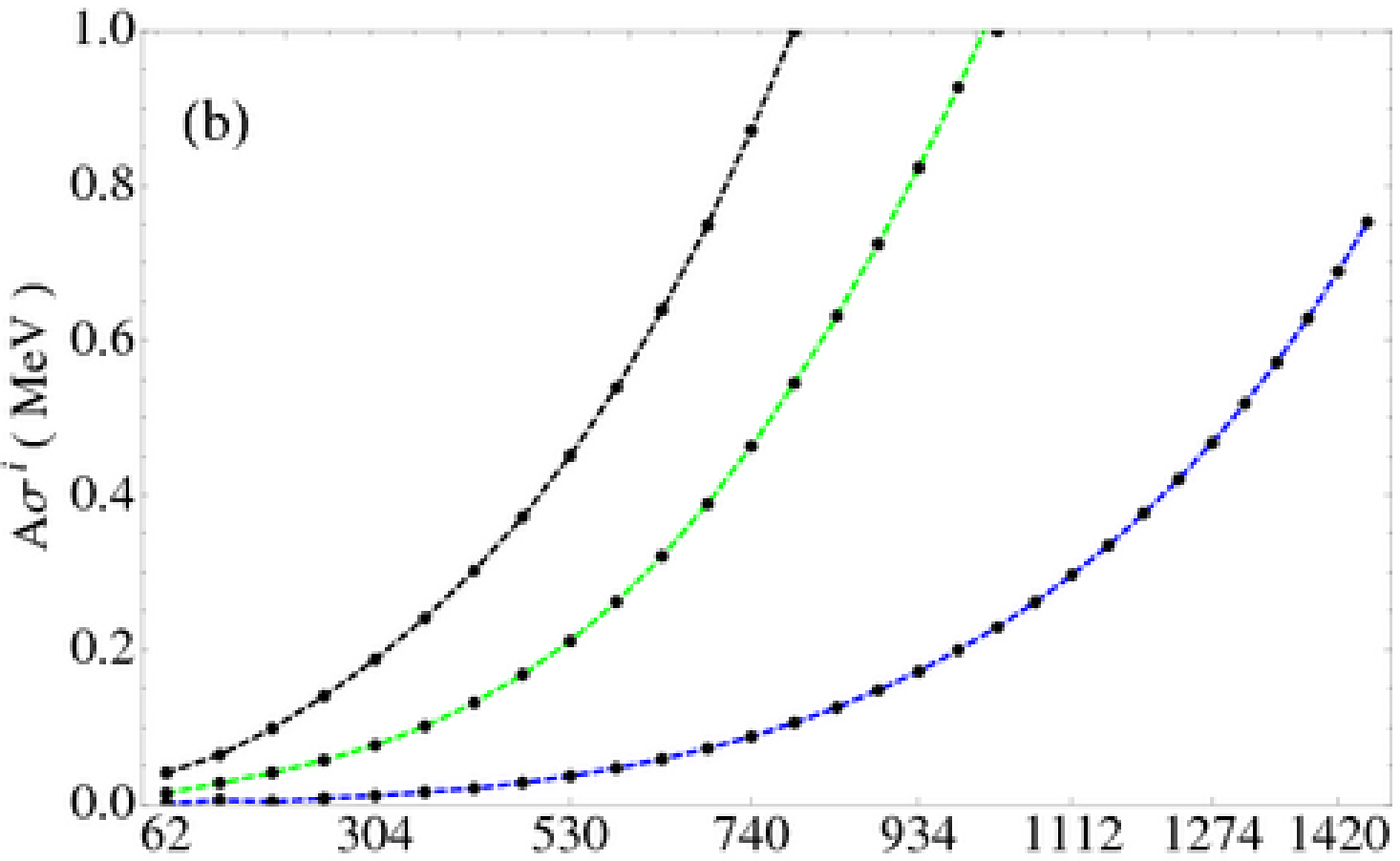}}
\caption{(a) Set of nuclei whose masses are predicted in each iteration for the long range test. (b) $A\sigma^{i}$ deviation (in MeV) for the mass prediction in the long range test using the masses predicted by different models, LDM(blue dotted),  LDMM(green dotted) and DZ(black).}
\label{gkbes}
\end{center}
\end{figure}

 Figure \ref{iterlot} shows the nuclei predicted on each iteration for this test. Note that in each iteration only nuclei with N-Z=constant are predicted. In this way, the distance to the known region and the number of iterations are the same and it is possible to analyze how the error grows as a function of this two variables.  We calculate the  accumulated root mean square deviation $A\sigma^{i}$ between the different models and the Garvey-Kelson prediction, for each iteration.  Figure \ref{rmsac} shows $A\sigma^{i}$ for these three models. We observe that for each of the three sets of data used, the rms grows systematically, the Duflo-Zucker  model being the one for which the error grows faster.\\

\begin{figure}[h]
\begin{center}
\qquad

\subfloat{\label{errldm}\includegraphics[width=8cm]{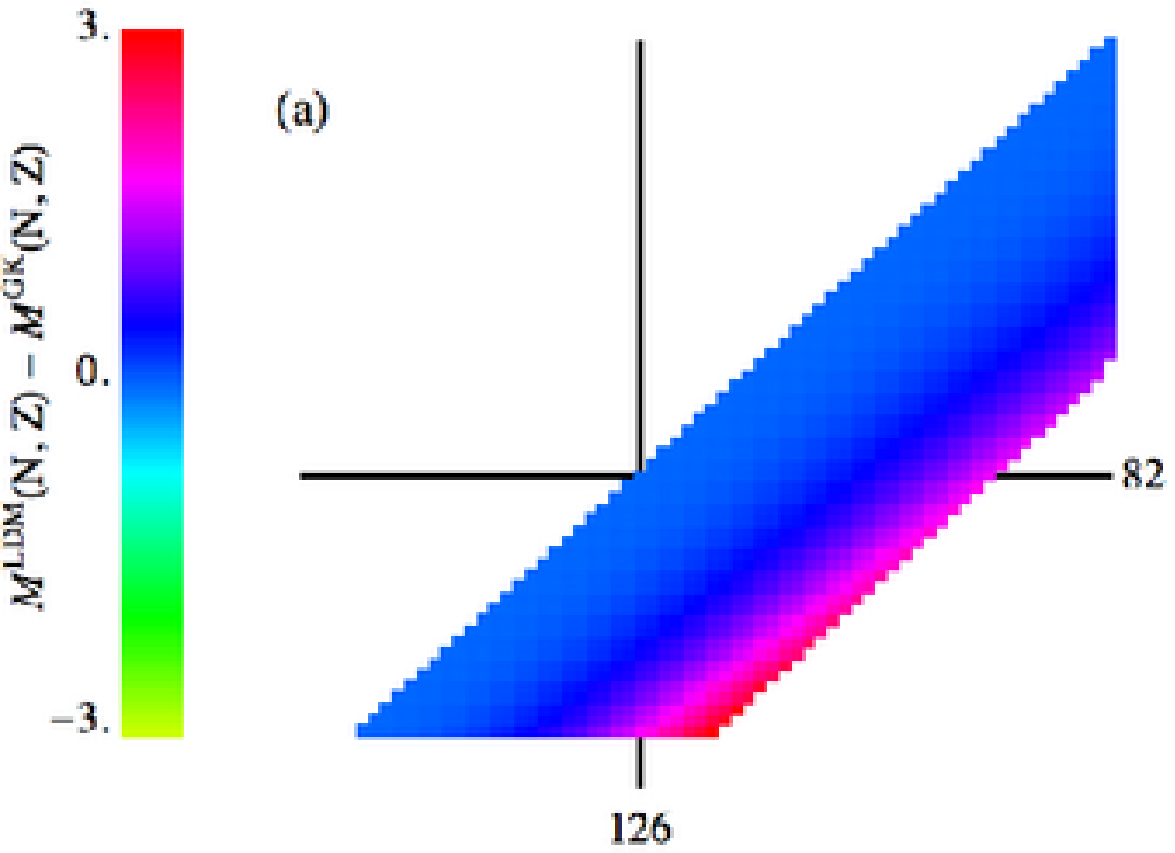}}\qquad
\subfloat{\label{inpldm}\includegraphics[width=8cm]{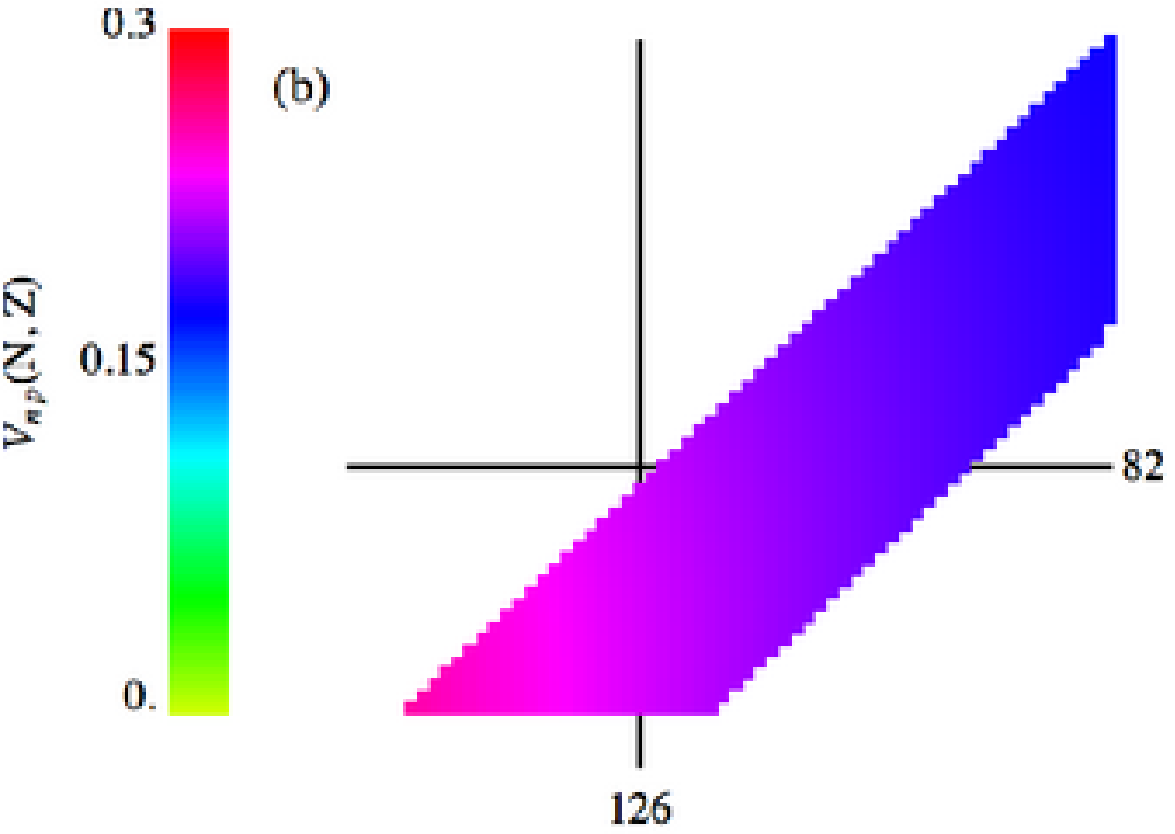}}
\caption{(a) Errors for the Garvey-Kelson iterative process using the predictions of LDM. (b) Average interaction of the last protons with the last neutrons $\delta V_{np}$. for LDM.}
\label{ldm}
\end{center}
\end{figure}

\begin{figure}[h]
\begin{center}
\qquad

\subfloat{\label{errldmm}\includegraphics[width=8cm]{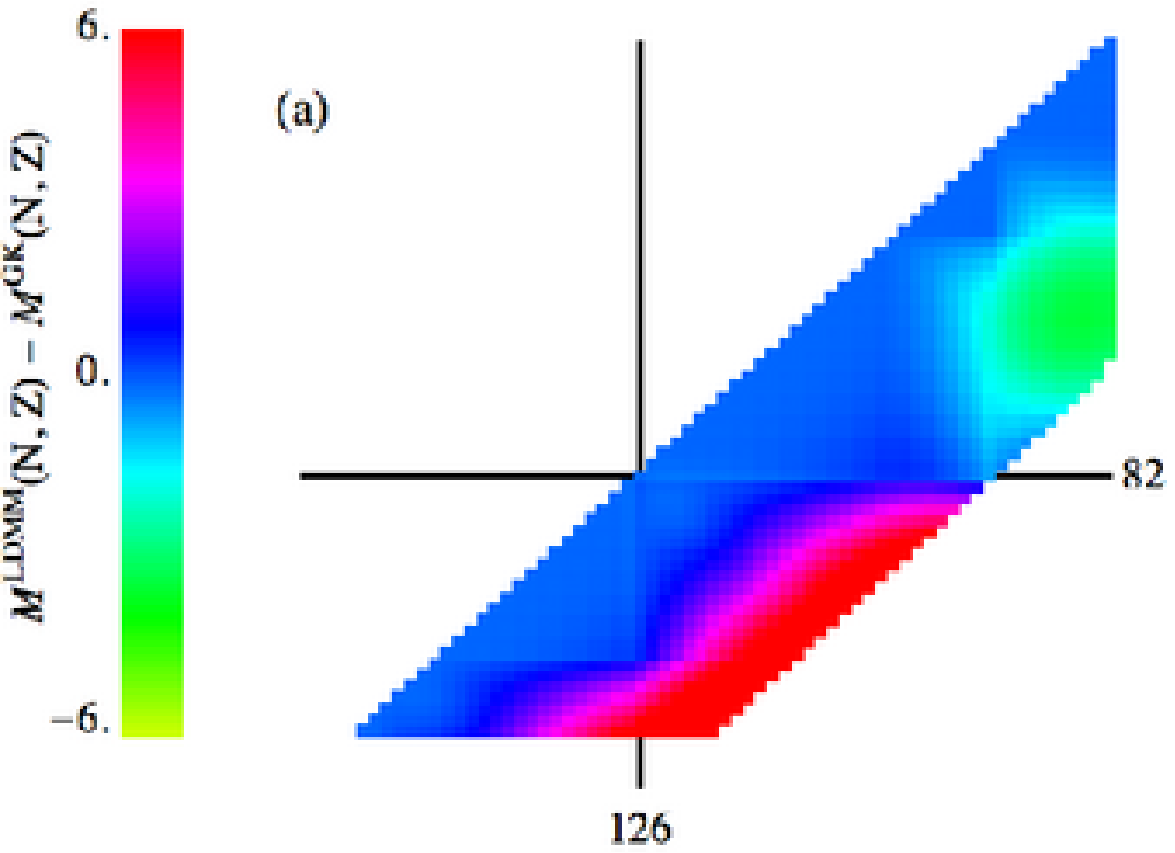}}\qquad
\subfloat{\label{inpldmm}\includegraphics[width=8cm]{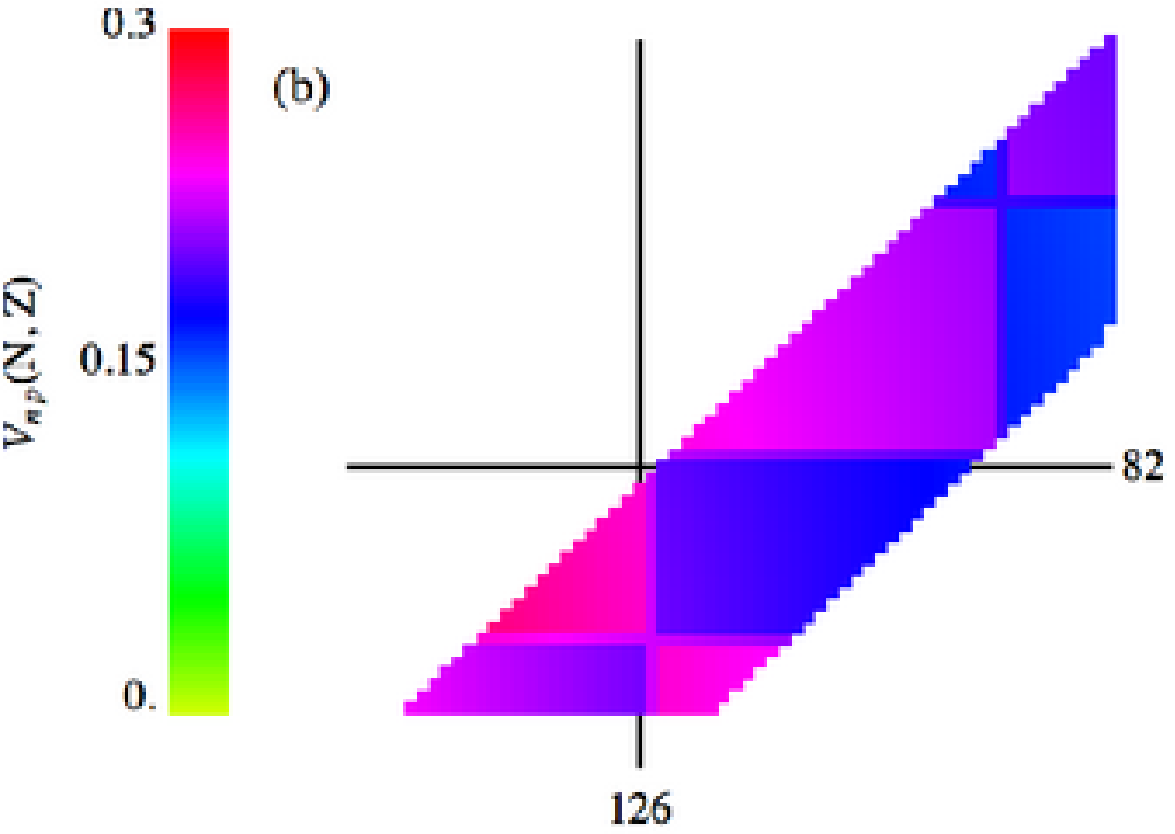}}
\caption{(a) Errors for the Garvey-Kelson iterative process using the predictions of LDMM. (b) Average interaction of the last protons with the last neutrons $\delta V_{np}$. for LDMM.}
\label{ldmm}
\end{center}
\end{figure}

\begin{figure}[h]
\begin{center}
\qquad

\subfloat{\label{errdz}\includegraphics[width=8cm]{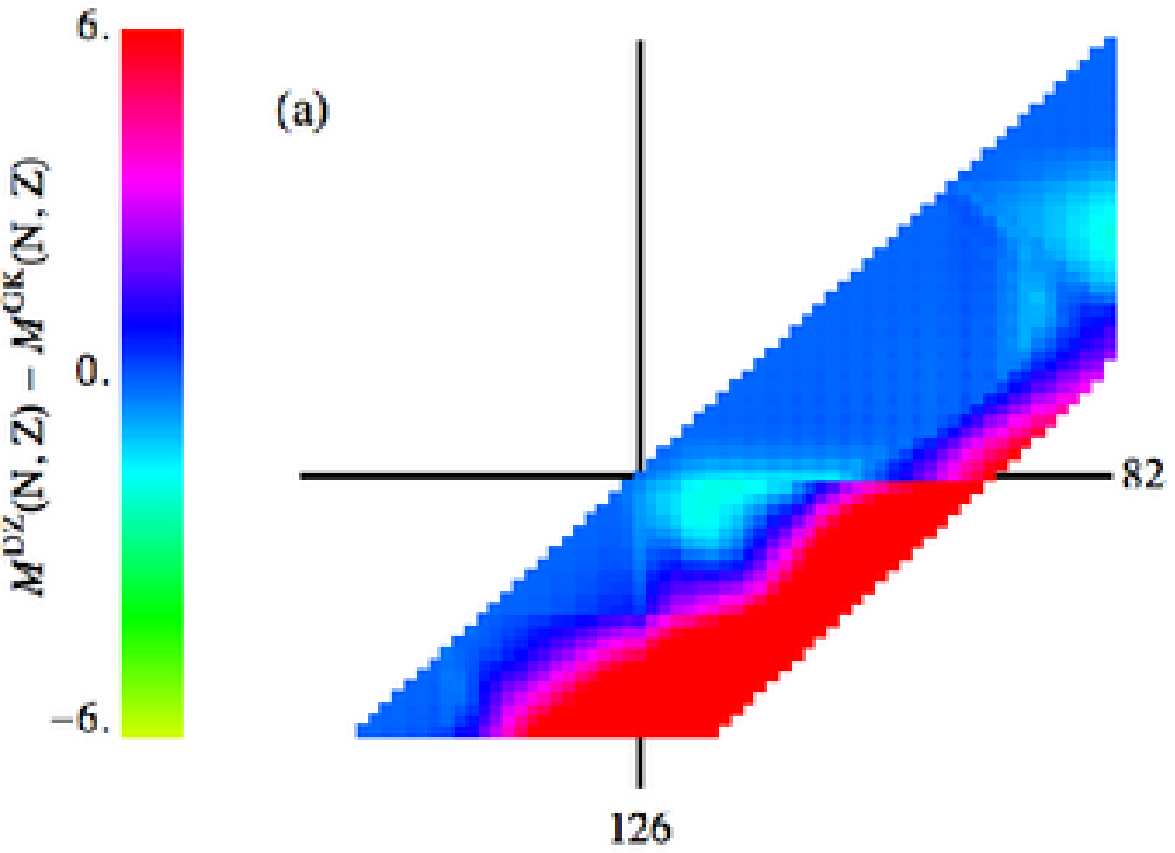}}\qquad
\subfloat{\label{inpdz}\includegraphics[width=8cm]{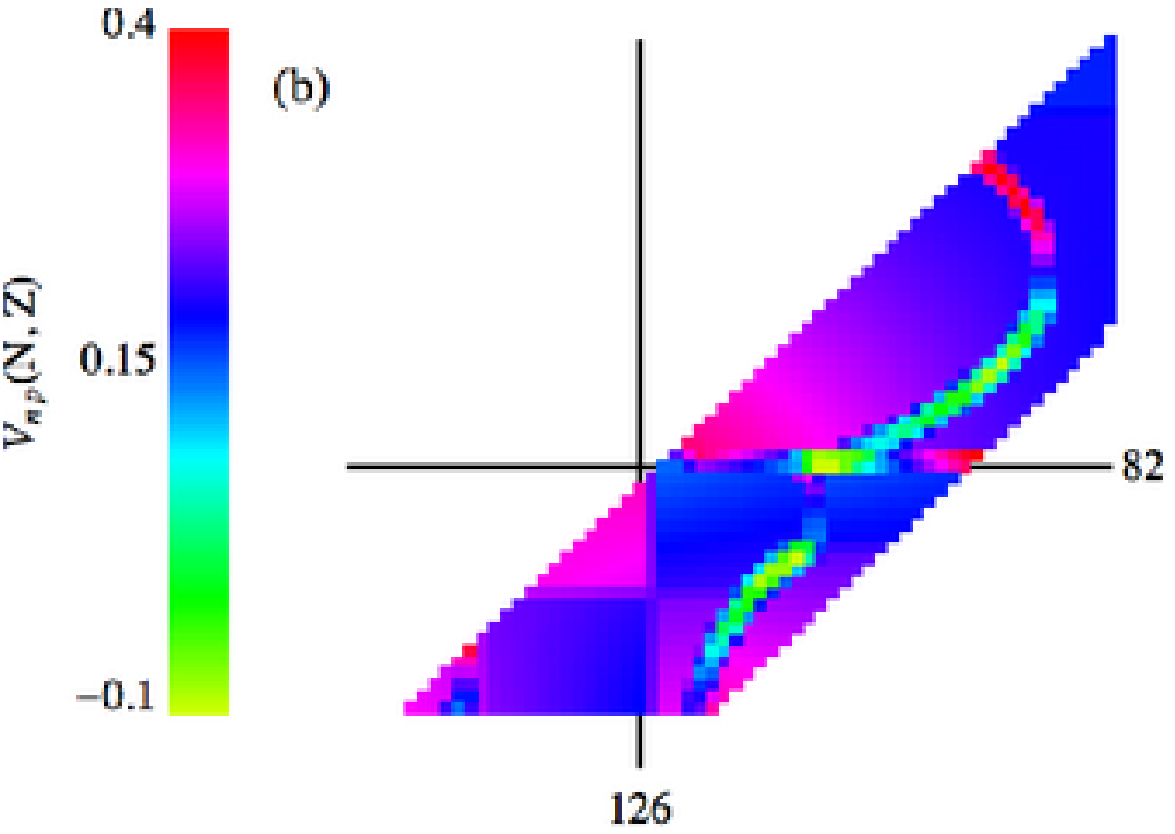}}

\caption{(a) Errors for the Garvey-Kelson iterative process using the predictions of DZ. (b) Average interaction of the last protons with the last neutrons $\delta V_{np}$. for DZ}
\label{dz}
\end{center}
\end{figure}

Figures \ref{errldm}, \ref{errldmm} and \ref{errdz} show the distribution of the error, $M^{model}(N,Z)-M^{GK}(N,Z)$, for LDM, LDMM and DZ sets of data respectively. It is remarkable that these errors are systematic and well localized. Taking into account that the three models satisfy the Garvey-Kelson relations remarkably well, we may assume that the systematic error is largely due to the iterative procedure.\\

One of the main assumptions of the Gravey-Kelson relations is that the residual interaction between nucleons vary smoothly with $N+Z$ and $N-Z$.  In order for the different masses to cancel, the Garvey-Kelson relations must be closely related to the interaction energy of the last $i$ neutrons and the last $j$ protons in a nucleus 

\begin{equation}
 \begin{array}{c}
\epsilon_{in-jp}(N+j,Z+i)=B(N+i,Z+j)-B(N,Z+j)-B(N+i,Z)+B(N,Z)
 \label{einjz}
 \end{array}
\end{equation}

Especially for $\epsilon_{1n-1p}$, $\epsilon_{1n-2p}$, $\epsilon_{2n-1p}$ and $\epsilon_{2n-2p}$, it is even possible to derive the Garvey-Kelson relations from assumptions about this quantities\cite{Zao01}. In fact, it can be shown that one of the relations assumes that $\epsilon_{1n-1p}$ is independent of $N+Z$ and the other assumes that $\epsilon_{1n-1p}$ is independent of $N-Z$. It has been shown\cite{Jan74} that this assumption is not followed by the experimental masses and it produces systematic errors in the predictions obtained with the Garvey-Kelson relations. If $\epsilon_{1n-1p}$ does not vary smoothly, then on each iteration of the Garvey-Kelson process the residual interactions do not cancel completely and a systematic error accumulates. Similar behavior has been observed previously in the average interaction of the last protons with the last neutrons $\delta V_{np}=\frac{1}{4}\epsilon_{2n-2p}$ and it has been shown that this quantity not only is not smooth but e
 xhibits very distinctive patterns\cite{Cak05}.\\

Figures \ref{inpldm}, \ref{inpldmm} and \ref{inpdz} show $\delta V_{np}$ calculated from data of each model. The LDM shows a very smooth variation of $\delta V_{np}$ with $N+Z$ and $N-Z$, this is the reason why the error of the iterative process grows very slowly. Comparing the error and $\delta V_{np}$ we can observe that both present the same trend and display some correlation. The variation of $\delta V_{np}$ in the LDMM is not so smooth, it has discontinuities at the shell closures and near the middle of them. This discontinuities are produced because the model has terms that take into account the number of valence nucleons and in this regions the valence particles change nature from "particles" to "holes" and vice versa. Figure \ref{errldmm} shows that this discontinuities, where $\delta V_{np}$ varies abruptly, are boundaries for the regions which display systematic errors.\\

The comparison between the errors and the $\delta V_{np}$ in the DZ case shows  more clearly the influence on the errors produced by the sudden variations of $\delta V_{np}$. The DZ model introduces a competition between two behaviors, for each shell,  in order to model the dependence of mass on deformation. The boundary between these two regimes creates a discontinuity that is enhanced in the $\delta V_{np}$. Figure \ref{inpdz} shows clearly one of this lines as well as the same discontinuities due to valence nucleons in the shell closures and in the midshells. Both kinds of discontinuities create boundaries for the regions with systematic errors.\\

\begin{figure}
\begin{center}
\subfloat{\label{vnpe}\includegraphics[width=8cm]{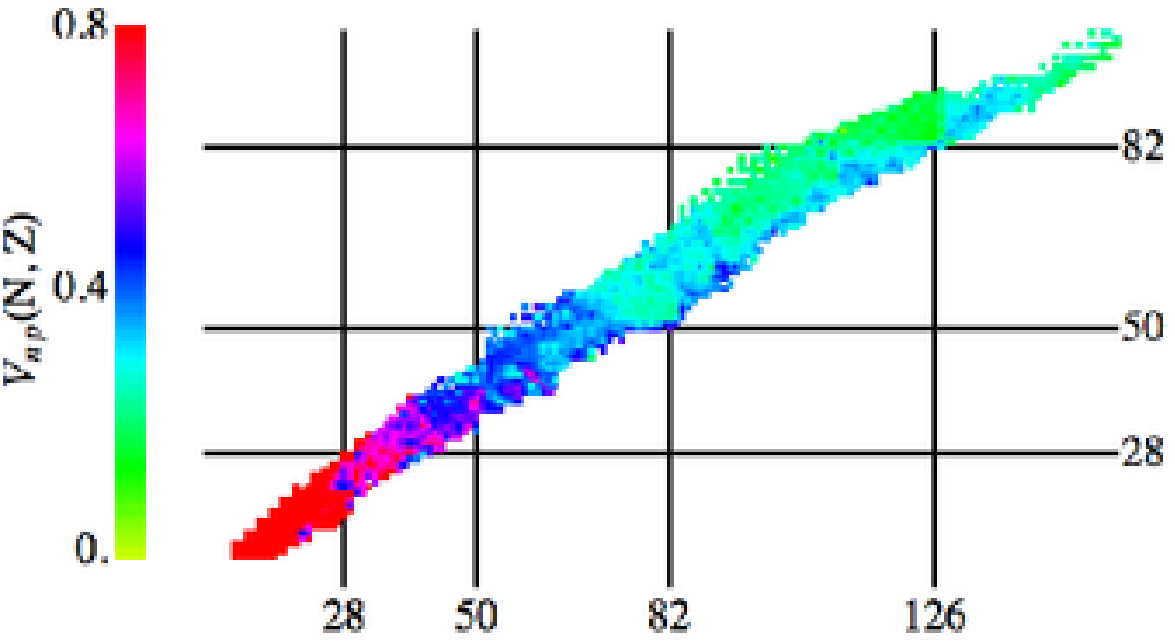}} 
\caption{Average interaction of the last protons with the last neutrons $\delta V_{np}$ for AME03 experimental data.}
\label{vnpexp}
\end{center}
\end{figure}

It is clear that there exists a correlation between the regions with systematic error and the residual neutron-proton interaction. At least two effects produce this correlation, the dependence of $\delta V_{np}$ with $N+Z$ and $N-Z$ and the discontinuities or abrupt variations in $\delta V_{np}$. The discontinuities form boundaries and the error correlates with them.\\These discontinuities have been observed in the behavior of the models and it is clear that the errors of the iterative process are correlated with their position. Figure \ref{vnpexp} and the analysis in \cite{Cak05} shows that the experimental data also presents discontinuities in $\delta V_{np}$. It is an important task to include these effects in order to improve the GK procedure.

\section{Conclusions.}
In conclusion, a systematic procedure to estimate nuclear masses close to the experimentally known region has been presented. This procedure uses the Garvey-Kelson relations through an iterative process. The predictive power of this procedure is measured as a function of the number of iterations in the process.\\

The predictions of the Garvey-Kelson iterative process have been compared with the predictions of several models: the Audi-Wapstra extrapolations, the Bethe-Weizsacker Liquid Drop Model, the Bethe-Weizsacker Liquid Drop Model  with schematic shell corrections in the number of valence nucleons and the Duflo-Zuker model, as a function of the distance to the known region. Special attention was given to the Audi-Wapstra extrapolation technique since, like the GK relations, it is based on the smoothness and the systematic trends of the mass surface in the N-Z plane, and because this method has the best estimates for nuclear masses in the short range.\\ 

In all cases the Garvey-Kelson mass predictions are better than those obtained from any other model for the first three or four iterations. This procedure seems particularly well suited to predict unknown masses of very unstable nuclei, starting from a set which can include the more than 400 nuclei whose masses were recently measured at the nuclear laboratories at GSI, ISOLDE, GANIL, MSU,  Jyvaskyla, Stockholm and FSU \cite{numass}.\\

A systematic study of the way the error grows as a function of the iteration and the distance to the known masses region shows that a correlation exists between the error and the residual neutron-proton interaction, produced mainly by the implicit assumption in the Garvey-Kelson relations that $V_{np}$ varies smoothly along the nuclear landscape. The discontinuities on $V_{np}$ produce systematic errors that grow with the number of iterations.\\

While the use of Garvey-Kelson relations is one of the simplest and more accurate methods to predict masses near the experimentally known region, it is necessary to develop strategies to increase the range of reliable predictions. These strategies must take into account the fact that $V_{np}$ does not always vary as smoothly as the Garvey-Kelson relations assume and thus it is necessary to design ways to correct this effect.\\ 

Thanks are due to C.Thibault for very useful comments. Partial financial support from Conacyt-Mexico and DGAPA-UNAM is acknowledged.  

\begin{widetext}
\begin{center}
\begin{table}[h]
\caption{$\sigma^{i}$ and $A\sigma^{i}$ deviation in MeV for the predictions on each iteration of the Garvey-Kelson iterative procedure. The first column shows the iteration number, in the second column the number of predicted nuclei up to that iteration, in parenthesis the nuclei predicted only on that iteration. A graphical comparison of the data is shown in Fig.~\ref{gaudi}.}
\label{audi}
\begin{tabular}{|c|c|c|c|c|c|c|c|c|c|c|c|c|c|}
\cline{1-2} \cline{4-5} \cline{7-8} \cline{10-11} \cline{13-14}
&nuclei&     &$\sigma^{i}$&$\sigma^{i}$&     &$\sigma_{mod}^{i}$&$\sigma_{mod}^{i}$&     &$A\sigma^{i}$&$A\sigma^{i}$&     &$A\sigma_{mod}^{i}$&$A\sigma_{mod}^{i}$\\
$i$&predicted&     &GK&Audi&     &GK&Audi&     &GK&Audi&     &GK&Audi\\
\cline{1-2} \cline{4-5} \cline{7-8} \cline{10-11} \cline{13-14}
1&94 (94)&     &0.3391&0.3160&     &0.1459&0.1092&     &0.3391&0.3160&     &0.1459&0.1092\\
\cline{1-2} \cline{4-5} \cline{7-8} \cline{10-11} \cline{13-14}
2&157 (63)&     &0.4039&0.4465&     &0.1596&0.1381&     &0.3742&0.3739&     &0.1334&0.1219\\
\cline{1-2} \cline{4-5} \cline{7-8} \cline{10-11} \cline{13-14}
3&198 (41)&     &0.3754&0.3513&     &0.3325&0.1479&     &0.3803&0.3693&     &0.1362&0.1275\\
\cline{1-2} \cline{4-5} \cline{7-8} \cline{10-11} \cline{13-14}
4&230 (32)&     &0.4685&0.1930&     &0.4702&0.1838&     &0.3838&0.3501&     &0.2539&0.1378\\
\cline{1-2} \cline{4-5} \cline{7-8} \cline{10-11} \cline{13-14}
5&256 (26)&     &0.7531&0.1168&     &0.7565&0.1060&     &0.4349&0.3340&     &0.3483&0.1346\\
\cline{1-2} \cline{4-5} \cline{7-8} \cline{10-11} \cline{13-14}
6&278 (22)&     &0.7700&0.1355&     &0.7801&0.0876&     &0.4681&0.3227&     &0.4012&0.1317\\
\cline{1-2} \cline{4-5} \cline{7-8} \cline{10-11} \cline{13-14}
7&291 (13)&     &0.5062&0.0973&     &0.5062&0.0929&     &0.4690&0.3161&     &0.4069&0.1297\\
\cline{1-2} \cline{4-5} \cline{7-8} \cline{10-11} \cline{13-14}
\end{tabular}
\end{table}
\end{center}
\end{widetext}

\end{document}